\documentclass[11pt,a4paper]{article}
\pdfoutput=1
\usepackage{jheppub}
\usepackage{bm}
\usepackage{booktabs}
\usepackage{microtype}
\usepackage{subcaption}
\usepackage{changes}
\newcommand{\ii}{\mathrm{i}}
\newcommand{\ee}{\mathrm{e}}
\newcommand{\dd}{\mathrm{d}}
\newcommand{\cN}{\mathcal N}
\newcommand{\dO}{\delta_{\Omega}}
\newcommand{\dt}{\delta_t}

\newcommand{\Order}{\mathcal O}

\title{Channel-Resolved High-Overtone Quasinormal Modes Decode Kasner Scaling inside Black Holes}

\author[a]{Zi-Qing Xiao,}
\author[a]{Jun Nian}
\author[b,c,d]{and Li Li}

\affiliation[a]{International Centre for Theoretical Physics Asia-Pacific (ICTP-AP),\\
University of Chinese Academy of Sciences, Beijing 100190, China}
\affiliation[b]{Institute of Theoretical Physics, Chinese Academy of Sciences,\\
Beijing 100190, China}
\affiliation[c]{School of Fundamental Physics and Mathematical Sciences,\\
Hangzhou Institute for Advanced Study, UCAS, Hangzhou 310024, China}
\affiliation[d]{School of Physical Sciences, University of Chinese Academy of Sciences, Beijing 100049, China}
\emailAdd{xiaoziqing24@mails.ucas.ac.cn, nianjun@ucas.ac.cn, liliphy@itp.ac.cn}

\abstract{
The high-overtone quasinormal-mode spectrum encodes the near-singularity
Kasner scaling of black-hole interiors.
We show that distinct structures in this spectrum independently determine
the temporal and spatial Kasner grades, allowing the corresponding metric
exponents to be reconstructed without imposing Kasner constraints or
invoking holography.
We demonstrate this independent reconstruction at the fixed Kasner point
of five-dimensional asymptotically flat Schwarzschild and planar
Schwarzschild--AdS$_5$, and then extend it to a continuously varying
Gibbons--Maeda black-hole family.
The reconstructed exponents satisfy the Kasner relation a posteriori,
providing a nontrivial geometric consistency check.
More generally, the high-overtone spectrum separates into global spectral
data and local algebraic corrections whose fractional powers are fixed by
the near-singularity Kasner geometry.
These fractional corrections therefore provide a direct spectral probe of
the local geometry behind the horizon.
}

\begin{document}
\setcounter{tocdepth}{1}
\maketitle
\flushbottom
\section{Introduction}
Black-hole quasinormal modes (QNMs) are the damped characteristic frequencies of black-hole perturbations and the basic observables of
ringdown spectroscopy \cite{ReggeWheeler1957,BertiCardosoStarinets2009,BertiCardosoWill2006,
Abbott2016GW150914,IsiEtAl2019,GieslerEtAl2019}. Low-lying modes are primarily sensitive to the exterior geometry. Highly damped modes probe a qualitatively different regime: monodromy and complex-WKB analyses show that their quantization requires analytic continuation of the radial wave equation through its complex singularity and turning-point structure \cite{Motl2003,MotlNeitzke2003,AnderssonHowls2004,
CardosoNatarioSchiappa2004,NatarioSchiappa2004,
FestucciaLiu2009,BiggsMaldacena2023}. This raises a natural inverse problem: can QNMs, defined entirely by horizon and outer-boundary conditions, reconstruct local metric data near a spacelike singularity? In this paper we show that distinct high-overtone spectral channels independently encode the temporal and spatial Kasner scaling, allowing the corresponding metric exponents to be reconstructed without imposing a Kasner constraint.

Black-hole interiors can be accessed indirectly through a variety of observables, but with different geometric information content.
Light-ring probes are largely controlled by exterior null geometry \cite{CardosoMirandaBertiWitekZanchin2009,
DodelsonIossaKarlssonLupsascaZhiboedov2024}, while holographic one-point functions and thermal correlators can probe analytically
continued or integrated interior data \cite{GrinbergMaldacena2021,FidkowskiEtAl2004,
HartmanMaldacena2013,CeplakEtAl2024,
HorowitzLeungQueimadaZhao2024,Chakravarty2026,
AfkhamiJeddi2026,JiaRangamani2026Exact}. Complexity, thermal $a$-function, timelike entanglement and related reconstruction schemes can likewise become sensitive to the near-singularity geometry and, in some cases, to Kasner data \cite{AuzziEtAl2022,CaceresKunduPatraShashi2022,
JorstadMyersRuan2023,AreanJeongPedrazaQu2024,
CaceresMurciaPatraPedraza2024,AnegawaTamaoka2024,
HellerOriSerantes2025,LiYang2026KasnerTEE,
JafferisLamprou2022,GaoLamprou2022,DeBoerJafferisLamprou2022,
PrihadiEtAl2025Scrambling,An:2022lvo}. Recent work has also shown that asymptotic QNM spectra can retain information about the black-hole interior through analytically continued propagation in cavity setups~\cite{GrozdanovMovrinValachQNM2026}. In such cavity or
partially-reflecting-boundary setups, the leading asymptotic spacing is controlled by a bouncing geodesic and hence by an integrated interior time,
$\omega_n \sim 2\pi n/t_*$~\cite{GrozdanovMovrinValachQNM2026}. This is complementary to the present analysis: the leading spacing is a global propagation effect, whereas the fractional corrections studied here encode the local Kasner geometry near the singularity. The issue is therefore not simply whether the interior leaves a spectral imprint. We instead ask whether standard black-hole QNMs, defined only by horizon and outer-boundary conditions, can be inverted into local metric data near the singularity.

Kasner-like spacelike singularities and sequences of Kasner epochs arise in a broad class of black-hole interiors, although the terminal behavior can depend on the matter content and gravitational dynamics \cite{Kasner1921,BelinskyKhalatnikovLifshitz1970,DoroshkevichNovikov1978,FrenkelEtAl2020,
HartnollHorowitzKruthoffSantos2020,HartnollHorowitzKruthoffSantos2021,
CaiLiYang2021,DiasHorowitzSantos2021,VanDeMoortel2024,
CaiGeLiYang2022,HartnollNeogi2023,CaiDuanLiYang2024,
CaiDuanLiYang2025,CaceresPatraPedraza2024,
DeClerckHartnollSantos2024,LiSunYang2026,CaceresEtAl2026,
DuanLiLiYang2026}. Near a local Kasner regime, we write
\begin{equation}
 ds^2\simeq-d\tau^2+c_t\tau^{2p_t}dt^2
 +\sum_i c_i\tau^{2p_i}dx_i^2 ,
 \label{eq:kasner}
\end{equation}
where $p_t$ and $p_i$ characterize the local metric scaling. These exponents may obey additional relations fixed by the underlying gravitational and matter equations, but no such relation will be assumed in the reconstruction below.

The key point is that different components of the near-singularity Kasner
geometry imprint distinct fractional-power corrections on the large-overtone
quasinormal frequencies $\omega_n$. We consider a probe free scalar field of
mass $\mu$. We show that the mass and momentum responses scale as
\begin{equation}
    \Delta\omega_n^{(\mu)}
    \sim \mu^2 n^{-\delta_t},
    \qquad
    \Delta\omega_n^{(k_i)}
    \sim k_i^2 n^{-\delta_i},
\end{equation}
with $n$ the overtone number, and
\begin{equation}
    \delta_t=\frac{2}{1-p_t},
    \qquad
    \delta_i=\frac{2(1-p_i)}{1-p_t}.
\label{eq:generalgrade}
\end{equation}
The temporal and spatial Kasner exponents are then reconstructed
directly from the spectral grades,
\begin{equation}
    p_t=1-\frac{2}{\delta_t},
    \qquad
    p_i=1-\frac{\delta_i}{\delta_t}.
\label{eq:reconstruct}
\end{equation}
These fractional powers originate from the singular inner scaling of the wave
equation. A compact derivation of the common local-to-spectral transfer---from
the inner double scaling through Bessel perturbation theory and matched
finite-part subtraction to the QNM root shift---is given in
Appendix~\ref{app:framework}. Crucially, the temporal and spatial exponents are reconstructed
independently: no Kasner relation is imposed to infer one from the
other.  The mass response determines $\delta_t$, while the momentum
response determines $\delta_i$.  The Kasner relation, when applicable,
can therefore be tested only after the reconstruction, providing an
independent geometric consistency check.

\begin{figure}[t]
\centering

\begin{subfigure}[t]{\textwidth}
    \centering
    \includegraphics[width=0.98\textwidth]{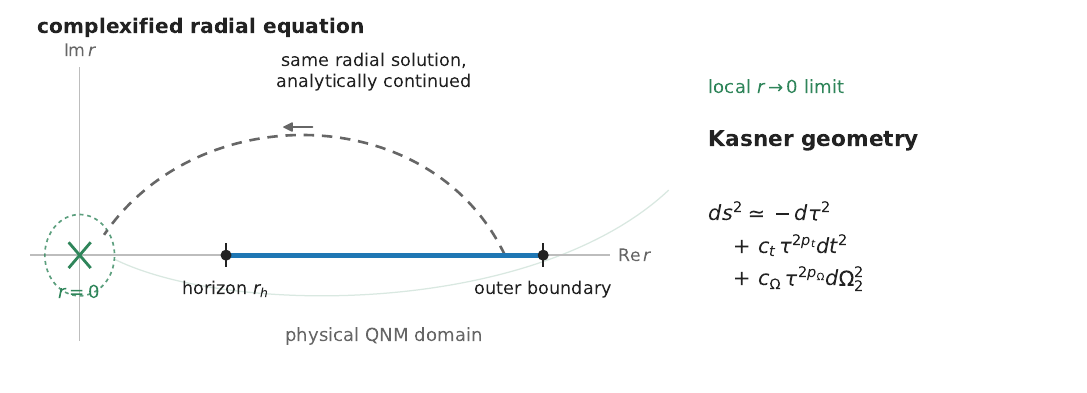}
    \caption{}
    \label{fig:mechanism-a}
\end{subfigure}

\vspace{1.5mm}

\begin{subfigure}[t]{\textwidth}
    \centering
    \includegraphics[width=0.98\textwidth]{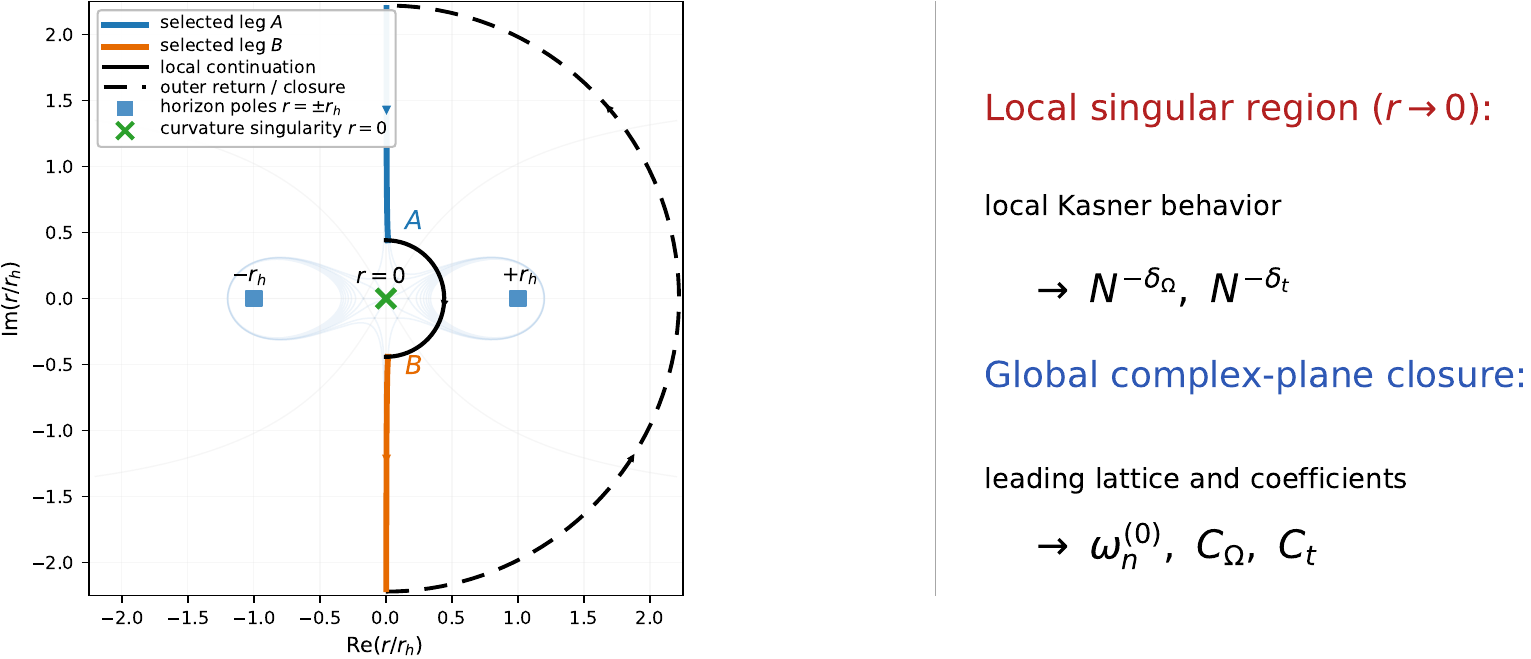}
    \caption{}
    \label{fig:mechanism-b}
\end{subfigure}

\caption{
Exterior-defined QNMs and the Kasner singularity.
\textbf{(a)} The physical radial interval is a real slice of the
complexified radial ODE; analytic continuation accesses the local
singular sector.
\textbf{(b)} In 5D Schwarzschild, the selected Stokes legs close
into the monodromy contour. Local singularity data fix the fractional
powers, while global closure fixes the leading lattice and complex
coefficients.
}
\label{fig:mechanism}
\end{figure}

The only nonlocal question is whether these local powers survive the full horizon-to-exterior connection problem.  Fig.~\ref{fig:mechanism} summarizes the mechanism: the physical QNM interval and the near-singularity sector belong to the same complexified radial ODE, so analytic continuation can transmit local Kasner scaling into an exterior-defined spectrum. We test this local/global separation in three stages: asymptotically flat Schwarzschild, Schwarzschild--AdS$_5$ with a different outer boundary condition, and a Gibbons--Maeda family with continuously varying Kasner exponents.

For five-dimensional (5D) asymptotically flat Schwarzschild and planar Schwarzschild--AdS$_5$, the local Kasner point is
$(p_t,p_i)=(-1/2,1/2)$, corresponding to $(\delta_t,\delta_i)=(4/3,2/3)$.  In both cases the temporal and spatial grades are extracted independently from distinct structures in the high-overtone spectrum, so that both Kasner exponents can be reconstructed without imposing a Kasner relation.  The asymptotically flat example establishes this reconstruction with standard outgoing QNM boundary conditions, while the AdS example shows that it survives a change of the outer boundary condition and of the global connection problem.  The Gibbons--Maeda family then extends the same reconstruction across a continuously varying family of Kasner interiors, with the raw spectrum providing an additional independent route to $p_\Omega$.  The Kasner relation is imposed nowhere in the reconstruction and is tested only afterwards as an independent geometric consistency check.

Our results separate the high-overtone spectrum into two qualitatively different structures: fractional-power corrections fixed by the local near-singularity scaling, and global contributions controlled by propagation in the complexified radial geometry.  The former admit a direct geometric inversion and determine the Kasner exponents, whereas the latter fix the leading spectral structure. The high-overtone spectrum therefore admits a separation between local
geometric data and global propagation effects. The paper is organized as follows. In Sec.~\ref{sec:AF} we establish independent reconstruction of the temporal and angular Kasner exponents for five-dimensional asymptotically flat Schwarzschild.  In Sec.~\ref{sec:AdS} we show that the same fixed-point reconstruction persists after changing the outer boundary condition to AdS normalizability.  In Sec.~\ref{sec:GM} we show that the same channel-resolved decoding extends across a continuously varying Gibbons--Maeda family. We discuss implications and future directions in Sec.~\ref{sec:discussion}.

\section{Kasner scaling from 5D Schwarzschild QNMs}

\label{sec:AF}
We begin with the 5D asymptotically flat
Schwarzschild black hole as a clean proof of principle. Its metric is
\begin{equation}
ds^2=-f(r)\,dt^2+\frac{dr^2}{f(r)}+r^2 d\Omega_3^2,
\qquad
f(r)=1-\frac{r_h^2}{r^2},
\label{eq:AF5-metric}
\end{equation}
where $r_h$ is the event-horizon radius and $d\Omega_3^2$
is the metric on the unit three-sphere. The choice is motivated by
two facts. First, the asymptotically flat black hole has the standard
outgoing QNM boundary condition at infinity, so the spectrum is defined
without introducing a cavity or a holographic dictionary. Second, the 5D Schwarzschild and planar Schwarzschild--AdS$_5$ share the same
local near-singularity Kasner geometry, allowing us to isolate the
effect of changing the outer boundary condition in
Sec.~\ref{sec:AdS}.
Near the spacelike singularity $r\to0$, $f(r)\simeq-r_h^2/r^2$. Introducing the interior proper time through
\begin{equation}
d\tau\simeq\frac{r}{r_h}\,dr,
\qquad
\tau\simeq\frac{r^2}{2r_h},
\end{equation}
the metric takes the local Kasner form
\begin{equation}
ds^2\simeq
-d\tau^2
+\frac{r_h}{2\tau}\,dt^2
+2r_h\tau\,d\Omega_3^2 .
\end{equation}
Hence the local Kasner exponents and the associated spectral grades are
\begin{equation}
 (p_t,p_{\Omega})=\left(-\frac12,\frac12\right),\qquad
 (\dt,\delta_{\Omega})=\left(\frac43,\frac23\right).
 \label{eq:schwpair}
\end{equation}
Here $\Omega$ labels the angular directions on $S^3$.  In the general
notation of equation \eqref{eq:generalgrade}, $p_i$ and $\delta_i$ refer to an arbitrary
spatial direction.  Spherical symmetry makes all three angular
directions equivalent, so that
\[
    p_i=p_\Omega,
    \qquad
    \delta_i=\delta_\Omega,
    \qquad i\in S^3 .
\]
Thus $p_\Omega$ is the single independent angular Kasner exponent of
the five-dimensional Schwarzschild interior.  The temporal and angular
grades, $\delta_t=4/3$ and $\delta_\Omega=2/3$, will be extracted
independently from the high-overtone spectrum.

To make the probe and the corresponding radial variable explicit, we
consider a minimally coupled scalar field $\Phi$ of mass $\mu$,
\begin{equation}
    \left(\Box-\mu^2\right)\Phi=0 .
    \label{eq:AF-KG}
\end{equation}
Separating the time and angular dependence as
\begin{equation}
    \Phi(t,r,\Omega_3)
    =
    e^{-i\omega t}\,
    Y_{\ell\mathbf m}(\Omega_3)\,
    r^{-3/2}\psi(r),
    \qquad
    \nabla_{S^3}^2Y_{\ell\mathbf m}
    =
    -L^2\,Y_{\ell\mathbf m},
    \qquad
    L^2=\ell(\ell+2),
    \label{eq:AF-separation}
\end{equation}
defines the Liouville-rescaled radial master field $\psi(r)$.
In terms of the tortoise coordinate,
\begin{equation}
    \frac{dr_*}{dr}=\frac{1}{f(r)},
\end{equation}
the radial equation takes the Schr\"odinger form
\begin{equation}
    \frac{d^2\psi}{dr_*^2}
    +
    \left[\omega^2-V(r)\right]\psi=0,
    \label{eq:AF-radial}
\end{equation}
with
\begin{equation}
    V(r)
    =
    f(r)
    \left[
        \frac{L^2}{r^2}
        +\mu^2
        +\frac{3f'(r)}{2r}
        +\frac{3f(r)}{4r^2}
    \right].
    \label{eq:AF-potential}
\end{equation}

For the 5D Schwarzschild geometry, the local tortoise coordinate
behaves near the singularity as
\begin{equation}
    r_* \simeq -\frac{r^3}{3r_h^2} + \cdots .
    \label{eq:AF-tortoise}
\end{equation}
Introducing the natural inner variable
\begin{equation}
    z=\omega r_*,
    \qquad
    \hat{\omega}\equiv\omega r_h,
    \qquad
    \hat{\mu}\equiv\mu r_h,
    \label{eq:AF-inner-variable}
\end{equation}
we regard the same radial master field as
$\psi(z)\equiv\psi[r(z)]$.  Using the near-singularity relation
$r_*\propto-r^3$, the radial equation \eqref{eq:AF-radial} reduces to
\begin{equation}
    \left[
    \frac{d^2}{dz^2}
    +1+\frac{1}{4z^2}
    +a_{\Omega}L^2\,\hat{\omega}^{-2/3}z^{-4/3}
    +a_t\hat{\mu}^{\,2}\hat{\omega}^{-4/3}z^{-2/3}
    +\cdots
    \right]\psi(z)=0 .
    \label{eq:AF-inner}
\end{equation}
Here $a_\Omega$ and $a_t$ are local coefficients. The two perturbations have a direct origin in the scalar potential \eqref{eq:AF-potential}:
the $L^2$ term is the angular-momentum response, whereas the $\hat{\mu}^2$ term is the mass response. Holding $z=\mathcal{O}(1)$ as $|\hat{\omega}|\to\infty$ therefore identifies the angular and mass responses at orders
\begin{equation}
    \hat{\omega}^{-2/3},
    \qquad
    \hat{\omega}^{-4/3},
\end{equation}
respectively. These are precisely the Kasner grades
$\delta_{\Omega}=2/3$ and $\delta_t=4/3$ obtained from the local
geometry in Eq.~\eqref{eq:schwpair}. In this sense, the fractional
powers are already fixed by the near-singularity scaling before the
global QNM boundary conditions are imposed.

The remaining question is whether these locally determined powers survive
after the full QNM boundary conditions are imposed. We address this using
the standard monodromy/complex-WKB strategy for highly damped QNMs
\cite{Motl2003,MotlNeitzke2003,AnderssonHowls2004,
CardosoNatarioSchiappa2004,NatarioSchiappa2004}.
The complex-plane continuation shown in Fig.~\ref{fig:mechanism-b}
connects the near-singularity solution to the horizon and to asymptotic
infinity. The geometry-specific global step is summarized in
Appendix~\ref{app:global-sch5}: it contains the regulated Bessel continuation,
the explicit outer-return audit, and the finite-part subtraction required in
the mass channel. The common matched-asymptotic argument is given in
Appendix~\ref{app:framework}. Requiring consistency after continuation around
the closed contour then yields the QNM condition

\begin{equation}
 \ee^{2\pi\hat\omega}
 =-3
 +m_{\Omega}L^2\,\hat\omega^{-2/3}
 +m_t\hat\mu^2\,\hat\omega^{-4/3}
 +\cdots,
 \label{eq:afqnm}
\end{equation}
where $m_\Omega$ and $m_t$ are complex coefficients determined by the
connection problem.

The constant $-3$ is fixed by the analytic continuation of the
leading near-singularity Bessel problem.  More importantly, the two
subleading terms retain exactly the powers predicted locally:
$\hat{\omega}^{-2/3}$ in the angular channel and
$\hat{\omega}^{-4/3}$ in the mass channel. Solving Eq.~\eqref{eq:afqnm} perturbatively at large overtone number
then gives
\begin{equation}
    \hat{\omega}_n
    =
    \frac{\log 3}{2\pi}
    -iN
    +C_\Omega^{\rm AF}L^2\,N^{-2/3}
    +C_t^{\rm AF}\hat{\mu}^{\,2}N^{-4/3}
    +\cdots ,
    \qquad
    N=n+\frac12 .
    \label{eq:afspectrum}
\end{equation}
The complex coefficients $C_\Omega^{\rm AF}$ and $C_t^{\rm AF}$ are fixed by the same connection problem. An independent Leaver--Nollert computation up to $n=500$ confirms both fractional powers and their complex coefficients, with relative agreement at the $10^{-4}$--$10^{-3}$ level \cite{Leaver1985,Nollert1993}. The continued-fraction formulation, implicit response derivatives, and convergence checks are collected in Appendix~\ref{app:num-sch5}.

Equations~\eqref{eq:afqnm} and \eqref{eq:afspectrum} provide the simplest realization of the reconstruction mechanism.  Although the QNMs are defined by boundary conditions at the horizon and at infinity, their high-overtone fine structure retains the fractional powers set by the near-singularity geometry.  The angular response determines $\delta_\Omega$, while the mass response determines $\delta_t$. Using equation~\eqref{eq:reconstruct}, these two independently measured grades reconstruct $p_\Omega$ and $p_t$, which may then be tested against the Kasner constraint.  In the next sections we show that the same mechanism persists for different asymptotic boundary conditions and for a continuously varying family of Kasner interiors.

\section{Changing the outer boundary condition}
\label{sec:AdS}

The asymptotically flat example shows that the local Kasner grades can
survive the full QNM connection problem. We now ask whether this is tied
to the outgoing boundary condition at infinity. To change the exterior
problem while keeping the same near-singularity geometry, we consider
planar Schwarzschild--AdS$_5$, for which the outer condition is instead
normalizability at the AdS boundary
\cite{HorowitzHubeny2000,HartnollKumar2005}.

We use the same outward-increasing radial convention as in
Sec.~\ref{sec:AF}: the curvature singularity is at $r\to0$, the event
horizon is at $r=r_h$, and the outer boundary is at $r\to\infty$.  With
the AdS radius denoted by $\ell_{\mathrm{AdS}}$, the metric is
\begin{equation}
    ds^2
    =
    -\frac{r^2}{\ell_{\mathrm{AdS}}^2}f(r)\,dt^2
    +\frac{\ell_{\mathrm{AdS}}^2}{r^2 f(r)}\,dr^2
    +\frac{r^2}{\ell_{\mathrm{AdS}}^2}d\mathbf{x}_3^2,
    \qquad
    f(r)=1-\frac{r_h^4}{r^4}.
    \label{eq:AdS-metric}
\end{equation}
The Hawking temperature and inverse temperature are
\begin{equation}
    T_H=\frac{r_h}{\pi\ell_{\mathrm{AdS}}^2},
    \qquad
    \beta_H=\frac{\pi\ell_{\mathrm{AdS}}^2}{r_h}.
    \label{eq:AdS-temperature}
\end{equation}
It is therefore natural to introduce the dimensionless frequency and
planar momentum
\begin{equation}
    \hat\omega
    \equiv
    \frac{\omega\ell_{\mathrm{AdS}}^2}{r_h}
    =\frac{\beta_H\omega}{\pi},
    \qquad
    \hat q
    \equiv
    \frac{q\ell_{\mathrm{AdS}}^2}{r_h}
    =\frac{\beta_H q}{\pi}.
    \label{eq:AdS-dimensionless}
\end{equation}
Near the spacelike singularity $r\to0$, one has
$f(r)\simeq-r_h^4/r^4$.  Introducing the interior proper time through
\begin{equation}
    d\tau\simeq
    \frac{\ell_{\mathrm{AdS}}\,r}{r_h^2}\,dr,
    \qquad
    \tau\simeq
    \frac{\ell_{\mathrm{AdS}}\,r^2}{2r_h^2},
    \label{eq:AdS-proper-time}
\end{equation}
the metric becomes
\begin{equation}
    ds^2
    \simeq
    -d\tau^2
    +\frac{r_h^2}{2\ell_{\mathrm{AdS}}\tau}\,dt^2
    +\frac{2r_h^2\tau}{\ell_{\mathrm{AdS}}^3}\,d\mathbf{x}_3^2.
    \label{eq:AdS-Kasner-metric}
\end{equation}
Hence
\begin{equation}
    (p_t,p_i)
    =
    \left(-\frac12,\frac12\right),
    \qquad
    (\delta_t,\delta_i)
    =
    \left(\frac43,\frac23\right),
    \label{eq:AdS-Kasner-grades}
\end{equation}
exactly as in the asymptotically flat benchmark.  Here $i$ labels any of
the three equivalent planar directions.

We use the same minimally coupled scalar probe as in Sec.~2, now
separating the planar directions as
\begin{equation}
    \Phi(t,r,\mathbf{x})
    =
    e^{-i\omega t+i\mathbf{q}\cdot\mathbf{x}}\phi(r),
    \qquad
    q\equiv|\mathbf{q}|.
\end{equation}
Near the AdS boundary, $r\to\infty$, the two independent radial
branches behave as
\begin{equation}
    \phi(r)
    \sim
    \phi_s\, r^{\Delta_\Phi-4}
    \left(1+\cdots\right)
    +
    \phi_v\, r^{-\Delta_\Phi}
    \left(1+\cdots\right),
    \label{eq:ads-boundary-falloff}
\end{equation}
where
\begin{equation}
    \Delta_\Phi(\Delta_\Phi-4)
    =
    \mu^2\ell_{\rm AdS}^2 .
\end{equation}
For standard quantization, $\phi_s$ and $\phi_v$ are respectively
the source and normalizable coefficients.  The quasinormal-mode
boundary condition at the AdS boundary is therefore
\begin{equation}
    \phi_s=0.
\end{equation}

  Thus the local probe is the same as in Sec.~\ref{sec:AF}, while
the outer boundary condition is genuinely different.

For the inner problem it is useful to define
\begin{equation}
    F(r)
    \equiv
    \frac{r^2}{\ell_{\mathrm{AdS}}^2}f(r),
    \qquad
    \frac{dr_*}{dr}=\frac{1}{F(r)}
    =\frac{\ell_{\mathrm{AdS}}^2}{r^2f(r)}.
    \label{eq:AdS-tortoise-def}
\end{equation}
After the Liouville rescaling
$\phi(r)=r^{-3/2}\psi(r)$, the radial equation takes the
Schr\"odinger form
\begin{equation}
    \frac{d^2\psi}{dr_*^2}
    +\left[\omega^2-V(r)\right]\psi=0,
    \label{eq:AdS-radial}
\end{equation}
with
\begin{equation}
    V(r)
    =
    F(r)
    \left[
        \frac{\ell_{\mathrm{AdS}}^2q^2}{r^2}
        +\mu^2
        +\frac{3F'(r)}{2r}
        +\frac{3F(r)}{4r^2}
    \right].
    \label{eq:AdS-potential}
\end{equation}
Let
\begin{equation}
    s\equiv r_*-r_*^{(S)},
    \label{eq:AdS-singular-tortoise}
\end{equation}
where $r_*^{(S)}$ is the limiting tortoise coordinate at the
singularity.  Equation~\eqref{eq:AdS-tortoise-def} gives
\begin{equation}
    s
    \simeq
    -\frac{\ell_{\mathrm{AdS}}^2}{3r_h^4}\,r^3,
    \qquad r\to0.
    \label{eq:AdS-tortoise-scaling}
\end{equation}
Thus the same local cubic relation $s\propto r^3$ that appeared in the
asymptotically flat example is present here as well.

Introducing the inner variable
\begin{equation}
    z=\omega s,
    \label{eq:AdS-inner-variable}
\end{equation}
and holding $z=\mathcal O(1)$ as $|\hat\omega|\to\infty$, the
near-singularity equation has the schematic graded form
\begin{equation}
    \left[
        \frac{d^2}{dz^2}
        +1+\frac{1}{4z^2}
        +a_i\hat q^{\,2}\hat\omega^{-2/3}z^{-4/3}
        +a_t\hat\omega^{-4/3}z^{-2/3}
        +\cdots
    \right]\psi(z)=0.
    \label{eq:AdS-inner}
\end{equation}
Here $a_i$ and $a_t$ are local coefficients; $a_t$ can depend on the
scalar mass through the dimensionless combination
$\mu^2\ell_{\mathrm{AdS}}^2$.  The momentum response therefore appears
at order $\hat\omega^{-2/3}$, while the next $q$-independent local
correction appears at order $\hat\omega^{-4/3}$.  These are precisely
the spatial and temporal Kasner grades in
Eq.~\eqref{eq:AdS-Kasner-grades}.

What changes is the global connection problem.  Unlike the
asymptotically flat case, the relevant continuation is not closed by an
infinity-to-horizon monodromy contour.  Instead, the solution selected
at the horizon is analytically continued to the AdS boundary and matched
to the normalizable falloff.  The corresponding complex-plane transport
is summarized in Appendix~\ref{app:global-ads5}, including the
finite-frequency Stokes geometry and the open matching condition imposed
by AdS normalizability.

The resulting high-overtone spectrum can be written directly in terms
of the dimensionless variables in Eq.~\eqref{eq:AdS-dimensionless} as
\begin{equation}
    \hat\omega_n
    =
    2(1-i)\cN
    +C_q^{\rm AdS}\hat q^2\cN^{-2/3}
    +C_t^{\rm AdS}\cN^{-4/3}
    +\cdots,
    \label{eq:adsspectrum}
\end{equation}
where
\begin{equation}
    \cN
    =
    n+\frac{\Delta_\Phi-3}{2}
    +\frac{i\log2}{2\pi}.
    \label{eq:AdS-overtone-variable}
\end{equation}
Here $C_q^{\rm AdS}$ and $C_t^{\rm AdS}$ are dimensionless complex
coefficients fixed by the global connection problem.

Equation~\eqref{eq:adsspectrum} cleanly separates the local and global
information.  The leading spacing and the complex coefficients differ
from their asymptotically flat counterparts because the exterior
boundary condition has changed.  The fractional powers, however, remain
$2/3$ and $4/3$: the momentum-dependent response carries the spatial
grade $\delta_i=2/3$, while the next Kasner-sensitive correction carries
the temporal grade $\delta_t=4/3$.  Using Eq.~\eqref{eq:reconstruct},
these two independently determined grades reconstruct $p_i=1/2$ and
$p_t=-1/2$ without imposing a Kasner constraint.  Thus changing outgoing
asymptotics to AdS normalizability changes the global spectral data
without changing the local Kasner grades.

The two geometries nevertheless approach this common asymptotic
structure at rather different rates.  The origin of this difference is
global rather than local.  In the asymptotically flat problem, spatial
infinity is an irregular singular endpoint of the radial equation, and
the QNM condition selects an outgoing oscillatory branch through
WKB/Stokes continuation.  By contrast, the AdS boundary is a regular
singular endpoint, where normalizability is imposed directly on the
local Frobenius falloffs.  This difference in endpoint structure is
consistent with the stronger finite-overtone contamination observed in
the asymptotically flat problem, whereas the AdS spectrum reaches the
Kasner-controlled asymptotic hierarchy at substantially lower overtones.
The near-singularity geometry thus fixes which fractional powers appear,
while the global analytic structure controls how quickly they become
visible at finite overtone number.

This distinction is also reflected numerically.  Using an independent
Horowitz--Hubeny computation, we find that both fractional powers and
their complex coefficients agree with the analytic prediction to better
than $10^{-3}$ already by $n=40$ \cite{HorowitzHubeny2000}, whereas the
asymptotically flat extraction requires substantially higher overtones
before the same fractional hierarchy is cleanly resolved.  Technical
details of the Horowitz--Hubeny recurrence, the implicit response
extraction, and the associated convergence checks are collected in
Appendix~\ref{app:num-ads}.

Although the AdS spectrum also admits the usual holographic
interpretation, no holographic dictionary is needed here: only the QNM
boundary condition and the analytic connection problem enter.  The AdS
example therefore shows that the Kasner-sensitive fractional powers
persist under a change of the outer boundary condition, even though the
rate at which they emerge from the finite-overtone spectrum is itself
controlled by the global problem.

\section{Continuously varying Kasner scaling in the Gibbons--Maeda family}
\label{sec:GM}

To move beyond a fixed Kasner point, we now consider a family of
backgrounds whose near-singularity geometry varies continuously.  We use
the four-dimensional asymptotically flat Gibbons--Maeda (GM) family of
charged black holes in Einstein--Maxwell--dilaton theory~\cite{GibbonsMaeda1988}.
Writing the conventional dilaton coupling as $a$, we use $A\equiv a^2$
throughout.  The Garfinkle--Horowitz--Strominger black hole arising in
low-energy string theory corresponds to the special value $a=1$ ($A=1$)
~\cite{GarfinkleHorowitzStrominger1991}.  Here we consider $A>1$ and use
its continuous variation to scan a family of distinct Kasner interiors.

In the Einstein frame, let $r_+$ denote the event-horizon radius and
write the curvature-singularity radius as $r_-=b r_+$, with $0<b<1$.
The metric is
\begin{equation}
    ds^2
    =
    -\Delta(r)\,dt^2
    +\frac{dr^2}{\Delta(r)}
    +R(r)^2\,d\Omega_2^2 ,
    \label{eq:gmmetric}
\end{equation}
with
\begin{equation}
\begin{aligned}
    \Delta(r)
    &=
    f_+(r)\,f_-(r)^{\frac{1-A}{1+A}},
    \qquad
    R(r)
    =
    r\,f_-(r)^{\frac{A}{1+A}},
    \\
    f_+(r)
    &=
    1-\frac{r_+}{r},
    \qquad
    f_-(r)
    =
    1-\frac{r_-}{r}
    =
    1-\frac{b r_+}{r}.
\end{aligned}
\label{eq:gmmetricdefs}
\end{equation}
Here $R(r)$ is the areal radius of the symmetry two-spheres.  To make
contact with the convention used in the calculations below, we introduce
horizon units
\begin{equation}
    \rho\equiv\frac{r}{r_+},
    \qquad
    \hat t\equiv\frac{t}{r_+},
    \qquad
    \hat\omega\equiv\omega r_+,
    \qquad
    \hat\mu\equiv\mu r_+,
    \qquad
    \hat\beta_H\equiv\frac{\beta_H}{r_+}.
    \label{eq:gm-horizon-units}
\end{equation}
Then $f_+=1-1/\rho$ and $f_-=1-b/\rho$, so the horizon and curvature
singularity are at $\rho=1$ and $\rho=b$, respectively.  For the
remainder of the GM analysis we work in these horizon units, set
$r_+=1$, and relabel $\rho\to r$ and $\hat t\to t$ for notational
simplicity.  The radial coordinate used below is therefore dimensionless.
We likewise measure the tortoise coordinate in units of $r_+$ and retain
hats on the dimensionless spectral parameters $\hat\omega$, $\hat\mu$,
and $\hat\beta_H$.

Near the spacelike singularity $r=b$, the proper time scales as
$\tau\propto(r-b)^{(3A+1)/[2(A+1)]}$, and the metric approaches a
Kasner geometry with
\begin{equation}
    p_t
    =
    \frac{1-A}{1+3A},
    \qquad
    p_{\Omega}
    =
    \frac{2A}{1+3A},
    \label{eq:gmkasner}
\end{equation}
with corresponding spectral grades
\begin{equation}
    \delta_{\Omega}
    =
    \frac{A+1}{2A},
    \qquad
    \delta_t
    =
    \frac{3A+1}{2A}
    =
    1+\delta_{\Omega}.
    \label{eq:gmdeltas}
\end{equation}
Thus, unlike the fixed Kasner point of the two previous examples, both
spectral grades vary continuously with $A$.  Their independence of $b$
already reflects the local nature of these powers: $b$ affects the
global spectral data, but not the Kasner grades themselves.

We probe the geometry with the same minimally coupled scalar field,
$(\Box-\mu^2)\Phi=0$.  In horizon units, separating the time and angular
dependence as
\begin{equation}
    \Phi(t,r,\Omega_2)
    =
    e^{-i\hat\omega t}
    Y_{\ell m}(\Omega_2)\,\phi(r),
    \qquad
    \nabla_{S^2}^2Y_{\ell m}
    =
    -L^2\,Y_{\ell m},
    \qquad
    L^2\equiv\ell(\ell+1),
    \label{eq:GM-scalar-ansatz}
\end{equation}
and defining the Liouville-rescaled master field
$\psi(r)\equiv R(r)\phi(r)$, the dimensionless tortoise coordinate
$dr_*/dr=\Delta^{-1}(r)$ puts the radial equation in Schr\"odinger form,
\begin{equation}
    \frac{d^2\psi}{dr_*^2}
    +\left[\hat\omega^2-V(r)\right]\psi=0.
    \label{eq:GM-radial}
\end{equation}
Here $V$ denotes the dimensionless horizon-unit potential and
$L^2=\ell(\ell+1)$ is the positive angular eigenvalue of
$-\nabla_{S^2}^2$, in contrast with $L^2=\ell(\ell+2)$ on $S^3$ in
Sec.~\ref{sec:AF}.  The physical inverse Hawking temperature is given by
$\beta_H$, while $\hat\beta_H=\beta_H/r_+$ is the dimensionless inverse temperature entering the spectral formulas below.

To expose the local spectral scaling, let
$s\equiv r_*-r_*^{(S)}$, where $r_*^{(S)}$ is the limiting tortoise
coordinate at $r=b$.  Near the singularity, the scalar potential takes
the form
\begin{equation}
    V(s)
    =
    -\frac{1}{4s^2}
    +
    \left(\lambda_1+L^2\lambda_{\Omega}\right)
    s^{-2+\delta_{\Omega}}
    +
    \hat\mu^2\lambda_t
    s^{-2+\delta_t}
    +\cdots ,
    \label{eq:gmlocal}
\end{equation}
where $\lambda_1$, $\lambda_{\Omega}$, and $\lambda_t$ are local
coefficients determined by the background.  With the natural inner
variable
\begin{equation}
    z=\hat\omega s,
\end{equation}
the geometric and angular contributions enter at order
$\hat\omega^{-\delta_{\Omega}}$, while the mass response enters at order
$\hat\mu^2\hat\omega^{-\delta_t}$.  Thus the same two grades that follow
from the metric in Eq.~\eqref{eq:gmdeltas} are already visible in the
local wave equation before the global QNM boundary conditions are imposed.

The remaining step is global.  The QNMs are ingoing at the event horizon
and outgoing at asymptotic infinity.  We impose these conditions using
the same monodromy/complex-WKB strategy as in Sec.~\ref{sec:AF}.  For the
rational values of $A$ used below, the analytic continuation can be
uniformized onto a single-valued complex $u$ plane.  The explicit
uniformization, endpoint periods, and the Stokes/winding audit over all
six production geometries are given in Appendix~\ref{app:global-gm}.
Requiring consistency with the horizon monodromy gives
\begin{equation}
\begin{aligned}
    e^{\hat\beta_H\hat\omega}
    ={}&
    -3
    +
    \left(m_1+m_{\Omega}L^2\right)
    \hat\omega^{-\delta_{\Omega}}
    +
    m_t\hat\mu^{\,2}
    \hat\omega^{-\delta_t}
    +\cdots
    +R_{\rm rem}(\hat\omega),
\end{aligned}
\label{eq:gmglobalqnm}
\end{equation}
where $m_1$, $m_{\Omega}$, and $m_t$ are complex coefficients fixed by
the connection problem.  The algebraic powers are precisely those
predicted by the local Kasner scaling.

The analytic continuation also contains a qualitatively different
contribution,
\begin{equation}
    R_{\rm rem}(\hat\omega)
    =
    -4
    \cos\!\left[
        \frac{\pi(A+1)}{2(A+3)}
    \right]
    \exp\!\left(
        2i\hat\omega\mathcal I_+
    \right),
    \label{eq:gmremote}
\end{equation}
where $\mathcal I_+$ is the dimensionless complex tortoise action, in
horizon units, associated with a remote singular point of the analytically
continued radial equation.  This point is not a second Lorentzian
singularity of the physical interior.  Importantly, $R_{\rm rem}$ is
exponential in $\hat\omega$, rather than algebraic in
$1/\hat\omega$, and can therefore generate finite-overtone oscillations
without changing the local fractional grades $\delta_{\Omega}$ and
$\delta_t$.  The GM example thus makes explicit that a single QNM spectrum
may contain both local algebraic information about the physical
singularity and distinct global information about the complexified radial
problem~\cite{DodelsonIossaKarlssonZhiboedov2024,
JiaRangamani2024,
ArnaudoIossaKarlssonWithers2026,
DodelsonIossaKarlsson2026,
GrozdanovMovrinValachCavity2026,
GrozdanovMovrinValachQNM2026}.

At leading order, the dimensionless GM monodromy condition gives
\begin{equation}
    e^{\hat\beta_H\hat\omega}=-3,
    \qquad
    \hat\omega_n^{(0)}
    =
    \frac{1}{\hat\beta_H}
    \left(\log 3-2\pi iN\right),
    \qquad
    N=n+\frac12 .
    \label{eq:gm-leading-lattice}
\end{equation}
We then determine the two Kasner grades from three spectrally distinct
observables:
\begin{equation}
\begin{aligned}
    \bigl(
        \hat\omega_n-\hat\omega_n^{(0)}
    \bigr)_{\rm alg}
    &\propto
    N^{-\delta_{\Omega}},
    \\
    \left.
    \frac{\partial\hat\omega_n}{\partial (L^2)}
    \right|_{L^2=0,\ \hat\mu=0}
    &\propto
    N^{-\delta_{\Omega}},
    \\
    \left.
    \frac{\partial\hat\omega_n}{\partial\hat\mu^2}
    \right|_{L^2=0,\ \hat\mu=0}
    &\propto
    N^{-\delta_t}.
\end{aligned}
\label{eq:gmroutes}
\end{equation}
The raw spectrum and the angular response therefore provide two
independent measurements of the same spatial grade $\delta_{\Omega}$,
while the mass response independently determines the temporal grade
$\delta_t$.  The remote contribution does not enter the leading
angular-response numerator, so agreement between the raw and angular
routes provides a nontrivial check that the extracted $\delta_{\Omega}$
is local.  Using the reconstruction map in Eq.~\eqref{eq:reconstruct},
these independently measured grades then determine $p_{\Omega}$ and
$p_t$ without imposing the Kasner relation $p_t+2p_{\Omega}=1$.

\begin{figure}[t]
\centering

\begin{subfigure}[t]{0.76\textwidth}
    \centering
    \includegraphics[width=\linewidth]
    {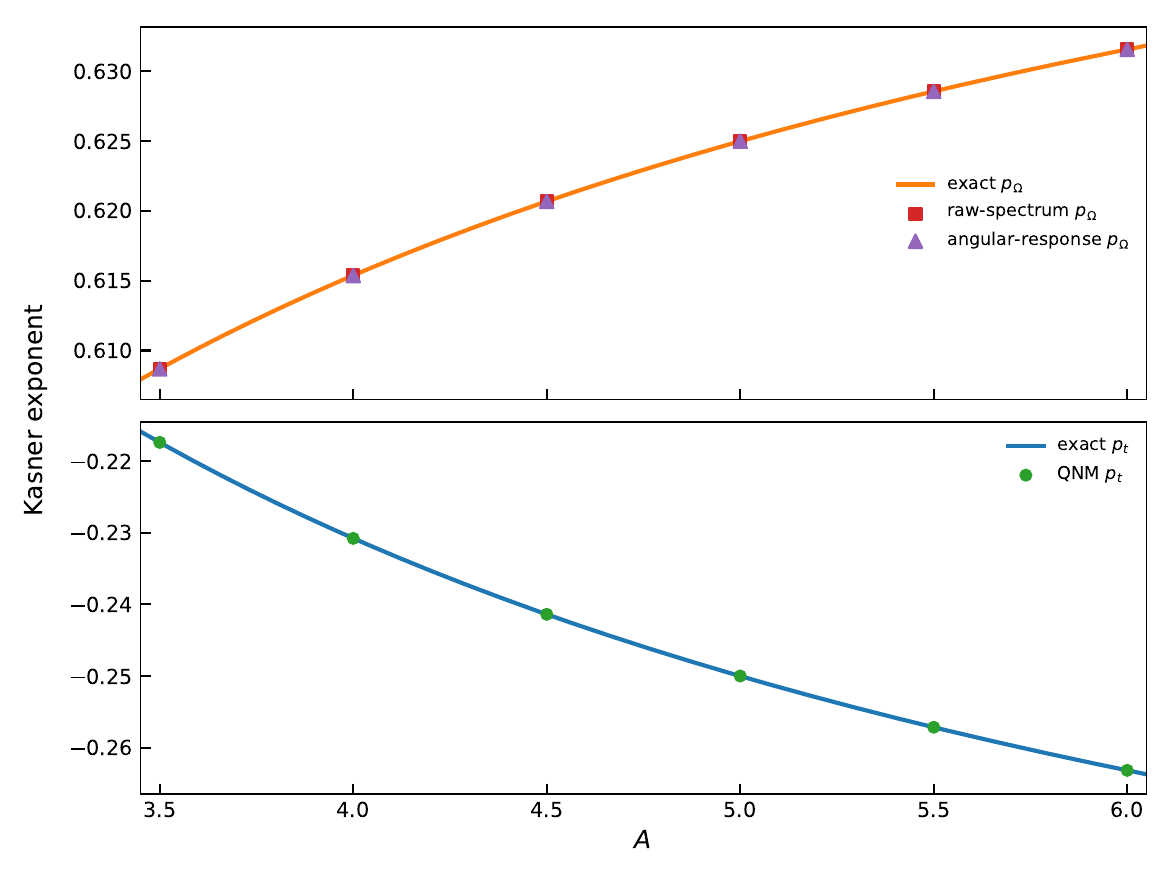}
    \caption{Reconstructed Kasner exponents.}
    \label{fig:gmresults-a}
\end{subfigure}

\vspace{1.5mm}

\begin{subfigure}[t]{0.48\textwidth}
    \centering
    \includegraphics[width=\linewidth]
    {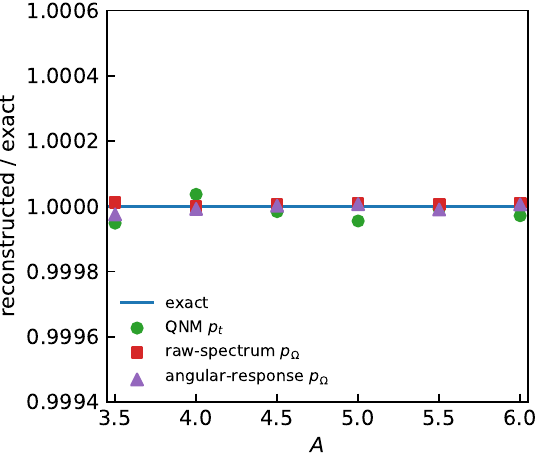}
    \caption{Reconstructed/exact ratios.}
    \label{fig:gmresults-b}
\end{subfigure}
\hfill
\begin{subfigure}[t]{0.48\textwidth}
    \centering
    \includegraphics[width=\linewidth]
    {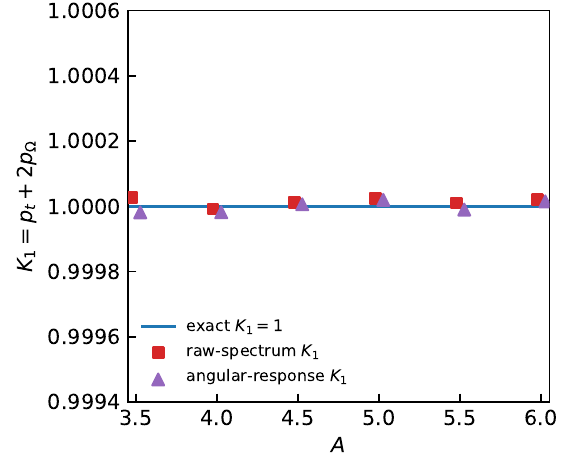}
    \caption{Held-out Kasner combination.}
    \label{fig:gmresults-c}
\end{subfigure}

\caption{
Channel-resolved extraction of the GM Kasner exponents.
Panel (a) shows the QNM-reconstructed temporal exponent $p_t$ and the
two independent reconstructions of $p_\Omega$, compared with the exact
GM metric curves.
Panel (b) shows the ratios of the reconstructed exponents to their exact
values.
Panel (c) shows the held-out Kasner combination
$K_1=p_t+2p_\Omega$, evaluated only after $p_t$ and $p_\Omega$ have
been independently reconstructed.
The temporal and angular spectral grades are extracted from distinct
response channels; neither an exact Kasner exponent nor any Kasner
constraint is imposed in the spectral fits.
}
\label{fig:gmresults}
\end{figure}

The QNMs are computed using an arbitrary-precision Evans formulation of
the exact finite-band recurrence.  The raw massless fits use consecutive
overtones in the common window $n=80,\ldots,120$, while the angular and
mass responses are fitted on consecutive overtone windows.  In each
production fit the analytic leading coefficient is imposed, but the
spectral exponent and subleading complex amplitudes remain free.
Neither an exact Kasner exponent nor a Kasner constraint therefore
enters the extraction; the Evans recurrence, response sensitivities,
fitting ansatzes, working-precision and cutoff checks, and less-assisted
fits are collected in Appendix~\ref{app:num-gm}.  Across the six
geometries $A=3.5,\,4,\,4.5,\,5,\,5.5,\,6$, the maximum deviations from
the exact grades are $6.5\times10^{-6}$ for the raw spatial route,
$3.1\times10^{-5}$ for the angular response, and $1.6\times10^{-5}$ for
the temporal response.  Substituting these measured grades into
Eq.~\eqref{eq:reconstruct} produces the reconstructed metric exponents
shown in Fig.~\ref{fig:gmresults}(\subref{fig:gmresults-a}), which agree
with the exact GM curves at the $10^{-5}$ level;
Fig.~\ref{fig:gmresults}(\subref{fig:gmresults-b}) shows the corresponding
reconstructed-to-exact ratios.

Only after $p_t$ and $p_\Omega$ have been independently reconstructed do
we form the held-out combination
$K_1^{\rm QNM}\equiv p_t^{\rm QNM}+2p_\Omega^{\rm QNM}$.  As shown in
Fig.~\ref{fig:gmresults}(\subref{fig:gmresults-c}), $K_1^{\rm QNM}$
remains consistent with unity across the six-point scan, with maximum
deviations $2.7\times10^{-5}$ for the raw spatial route and
$2.0\times10^{-5}$ for the angular route.  The Kasner relation is
therefore a genuinely held-out geometric test rather than an input used
to complete the reconstruction.

\section{Conclusion and Discussion}
\label{sec:discussion}

The three examples establish a common mechanism by which high-overtone
QNMs encode local Kasner scaling. In both Schwarzschild benchmarks,
distinct structures in the high-overtone spectrum independently determine
the temporal and spatial Kasner grades, which can then be inverted to
recover the corresponding metric exponents without imposing a Kasner
constraint. Five-dimensional asymptotically flat Schwarzschild demonstrates
this reconstruction with the standard ingoing-at-horizon and
outgoing-at-infinity QNM boundary conditions, while
Schwarzschild--AdS$_5$ shows that the same local Kasner grades persist
after changing the outer boundary condition and the global complex-plane
connection problem. The Gibbons--Maeda family then extends the same
channel-resolved reconstruction from a fixed Kasner point to a
continuously varying family of interiors. There, the angular and mass
responses determine $\delta_\Omega$ and $\delta_t$ independently, while
the raw spectrum provides a second independent route to
$\delta_\Omega$. The corresponding metric exponents are reconstructed
throughout the family without imposing the Kasner relation, which is
instead tested only afterwards as a held-out geometric constraint.

This independence is important beyond the particular examples studied
here. Relations among Kasner exponents are dynamical consequences of
the underlying gravitational and matter equations rather than
kinematical properties of a local power-law metric. In particular,
higher-derivative corrections can modify the classical Kasner dynamics
and can even lead to near-singularity regimes beyond ordinary Kasner
epochs~\cite{DuanLiLiYang2026,Bueno:2024fzg}. Inferring one metric
exponent from another by imposing a classical Kasner relation would
then remove precisely the information needed to diagnose such
departures. An independent spectral determination of the individual
exponents instead allows any putative Kasner relation to be tested
only after the geometry has been inferred. The constraint-free
character of the reconstruction is therefore not merely a redundancy
check, but is essential if high-overtone QNMs are to remain useful as
probes when the near-singularity dynamics itself is modified.

A further lesson of the Gibbons--Maeda example is that the same
high-overtone spectrum can encode qualitatively different kinds of
interior information. The algebraic fractional-power corrections are
fixed by the local near-singularity scaling and admit the direct
geometric inversion developed here, whereas the additional exponential
contributions retain information about the global analytic continuation
of the radial problem. Disentangling these sectors suggests a broader
form of black-hole spectroscopy in which local near-singularity geometry
and global complexified structure are inferred from different asymptotic
features of the same spectrum.

We emphasize that QNMs are not the only probes of black-hole interiors.
Timelike entanglement, complexity, thermal $a$-function diagnostics,
thermal correlators, and modular reconstruction can also access
behind-horizon or Kasner data~\cite{
JorstadMyersRuan2023,
AreanJeongPedrazaQu2024,
LiYang2026KasnerTEE,
Chakravarty2026,
DeBoerJafferisLamprou2022,
An:2022lvo}.
The distinguishing feature of the high-overtone QNM spectrum is that
it is defined solely by the bulk wave equation together with horizon
and outer-boundary conditions, requires no holographic dictionary, and
can be inverted into local metric exponents through independent
spectral structures. This is what allows the high-overtone spectrum
to encode not only global spectral data but also quantitative
information about the local Kasner geometry behind the horizon.

Although our primary interest is in asymptotically flat black holes,
which provide the standard setting for gravitational-wave ringdown,
the AdS case discussed here has an independent motivation. Through the
AdS/CFT correspondence, AdS QNMs are poles of retarded correlation
functions in the dual strongly coupled field theory~\cite{Kovtun:2005ev}
and therefore correspond to excitations of the dual quantum system.
The same fractional-power corrections are consequently encoded in the
large-frequency pole structure of the dual retarded correlators. It
would be interesting to make this connection precise and to understand
how the channel-resolved encoding of Kasner scaling identified here is
realized in the dual field-theoretic description.

The present result should nevertheless be viewed as a structural
statement about the asymptotic QNM spectrum rather than a near-term
observational proposal, since the relevant modes lie deep in the
highly damped regime. Several extensions are therefore important.
Rotating black holes would test whether the channel-resolved
reconstruction survives in the presence of mode coupling and a more
complicated complex-plane structure. Moving toward lower overtones
will likewise require control over finite-overtone systematics and
spectral non-normality. Finally, more complicated interiors---including
multiple Kasner epochs, BKL/Mixmaster dynamics, and Cauchy or weak-null
singularities---provide a broader arena in which to ask whether distinct
local scaling regimes generate separable structures in the asymptotic
spectrum.

\acknowledgments
We would like to thank Zi-Hao Li and Run-Qiu Yang for many helpful comments and discussions. This work was supported in part by the National Natural Science Foundation of China under Grants No.~12375067, No.\,12525503, No.~12547104, No. 12588101, and No.\,12447101.

\paragraph{Note added.}
While this work was being completed, Ref.~\cite{HartnollZhiboedov2026} appeared, extracting
Kasner-sensitive non-analytic powers from high-overtone momentum
dispersion in asymptotically AdS black holes.  In that approach the
temporal exponent is fixed using the Kasner constraint, whereas here the
spatial and temporal spectral grades are extracted independently and the
constraint is used only as an a posteriori check.

\appendix

\section{Local power transfer and matched asymptotics}
\label{app:framework}

This appendix records only the common ingredients needed to connect the local Kasner scaling to the large-overtone QNM spectrum.  Geometry-specific global closure is deferred to Appendix~\ref{app:global}.

For the Kasner metric in Eq.~\eqref{eq:kasner}, the distinguished singular limit keeps
\begin{equation}
 z\propto \omega\tau^{1-p_t}=\Order(1),
 \qquad
 \tau\propto \left(\frac{z}{\omega}\right)^{1/(1-p_t)}.
 \label{eq:appA-scaling}
\end{equation}
After the Liouville rescaling, the momentum and probe-mass terms therefore enter the inner equation as
\begin{equation}
 k_i^2\,\omega^{-\delta_i}z^{-2+\delta_i},
 \qquad
 \mu^2\,\omega^{-\delta_t}z^{-2+\delta_t},
 \qquad
 \delta_i=\frac{2(1-p_i)}{1-p_t},\quad
 \delta_t=\frac{2}{1-p_t}.
 \label{eq:appA-grades}
\end{equation}
This is the local origin of Eq.~\eqref{eq:reconstruct}; no Kasner constraint is used.

In all three examples the leading singular problem is of Bessel type.  Schematically,
\begin{equation}
 \left[\partial_z^2+1+\frac{1-j^2}{4z^2}
 +\sum_a \lambda_a\omega^{-\delta_a}z^{-2+\delta_a}+\cdots\right]\psi=0,
 \label{eq:appA-inner}
\end{equation}
where the regulator $j$ is kept until the connection coefficients are formed.  A single variation-of-parameters insertion leaves the explicit factor $\omega^{-\delta_a}$ outside the dimensionless Bessel integral, while repeated insertions generate sums of grades.  Thus
\begin{equation}
 \mathcal C_{\rm sing}(\omega)
 =\mathcal C_0+\sum_a\omega^{-\delta_a}\mathcal C_a
 +\sum_{a,b}\omega^{-(\delta_a+\delta_b)}\mathcal C_{ab}+\cdots .
 \label{eq:appA-connection}
\end{equation}
When the Bessel integral is not absolutely convergent at large $z$, a matching cutoff $1\ll \mathcal R\ll|\omega|$ separates it as
\begin{equation}
 I_{\rm inner}(\mathcal R)=I_{\rm overlap}(\mathcal R)+I_{\rm conn}+o(1).
 \label{eq:appA-finitepart}
\end{equation}
The first term is reproduced by the outer WKB phase and is subtracted once; the cutoff-independent remainder $I_{\rm conn}$ is the physical connection coefficient.  Equivalently, it is represented by the analytically continued Weber--Schafheitlin moment
\begin{equation}
 I(\delta)\equiv {\rm FP}\!\int_0^\infty z^{\delta-1}J_0(z)^2\,\dd z
 =2^{\delta-1}\frac{\Gamma(1-\delta)\Gamma(\delta/2)}{\Gamma(1-\delta/2)^3}.
 \label{eq:appA-moment}
\end{equation}
The genuine derivative-generated WKB error is also harmless at the grades used here.  If $\delta q=\omega^{-\delta}\Order(z^{-2+\delta})$, the WKB residual gives
\begin{equation}
 \delta H_{\rm WKB}^{(\delta)}
 =\omega^{-\delta}\Order(\mathcal R^{-3+\delta}),
 \label{eq:appA-higherwkb}
\end{equation}
which vanishes in the overlap for all the metric-sensitive exponents appearing below.

Finally, the QNM condition may be written as an open spectral determinant
$\mathcal D_{\rm QNM}=W[\psi_H^{\rm in},\psi_O^{\rm phys}]$.  A nonsingular change of endpoint normalization multiplies $\mathcal D_{\rm QNM}$ but cannot move its zeros.  Transporting the horizon monodromy into the outer basis gives the equivalent closed-monodromy eigenline condition, so the open and closed formulations used below are exact reformulations of the same boundary-value problem.  If a simple leading root obeys
\begin{equation}
 \mathcal D(\omega)=\mathcal D_0(\omega)+\omega^{-\delta}\mathcal D_1(\omega)+\cdots,
 \qquad \mathcal D_0(\omega_n^{(0)})=0,
 \label{eq:appA-det}
\end{equation}
then ordinary root perturbation gives $\omega_n-\omega_n^{(0)}\propto (\omega_n^{(0)})^{-\delta}$.  This is the local-to-spectral power-transfer statement used throughout the paper.

\section{Global completion in the three geometries}
\label{app:global}

The local fractional powers are determined by the singular inner problem, but their realization in the QNM spectrum depends on the global boundary-value problem.  This appendix records the geometry-specific completion needed in the three examples used in the main text.

\subsection{Five-dimensional Schwarzschild}
\label{app:global-sch5}

With $y=r/r_h$, $\hat\omega=\omega r_h$, $\hat\mu=\mu r_h$ and $L^2=\ell(\ell+2)$,
\begin{equation}
 \frac{r_*-r_*^{(S)}}{r_h}=-\frac{y^3}{3}-\frac{y^5}{5}-\cdots,
 \qquad |y|\sim |\hat\omega|^{-1/3},
 \label{eq:appB-sch5-scaling}
\end{equation}
so the two probe-sensitive local grades are $L^2\hat\omega^{-2/3}$ and $\hat\mu^2\hat\omega^{-4/3}$.  The physical infinity-connected Stokes legs are joined by the clockwise continuation $\Delta\arg z=-3\pi$.  Regulating the Bessel index and imposing the transported outgoing condition gives
\begin{equation}
 R_0=-\frac{\sin(3\pi j/2)}{\sin(\pi j/2)}
 =-(1+2\cos\pi j)\xrightarrow{j\to0}-3.
 \label{eq:appB-sch5-minus3}
\end{equation}

The only possible contamination at the same grades comes from the outer return.  Expanding the leading WKB phase between equal-magnitude matching points gives
\begin{equation}
 \delta\Theta_1
 =\Order(\mathcal R^{-1})
 +L^2\hat\omega^{-2/3}\Order(\mathcal R^{-1/3})
 +\hat\mu^2\hat\omega^{-4/3}\Order(\mathcal R^{1/3}).
 \label{eq:appB-sch5-overlap}
\end{equation}
The first two terms vanish in the overlap; the last is precisely the overlap term subtracted from the mass-channel Bessel integral.  Genuine higher-WKB corrections vanish with negative powers of $\mathcal R$, and triangular Stokes jumps protect the same dominant coefficient on the right-half-plane return.  The closed contour is homologous to a single physical-horizon loop, yielding
\begin{equation}
 e^{2\pi\hat\omega}
 =-3+m_\Omega L^2\hat\omega^{-2/3}
 +m_t\hat\mu^2\hat\omega^{-4/3}+\cdots,
 \label{eq:appB-sch5-qnm}
\end{equation}
which is the closed-contour form of Eq.~\eqref{eq:afqnm}.

\subsection{Planar Schwarzschild AdS$_5$}
\label{app:global-ads5}

Using the standard AdS radial coordinate $\rho=1/r$ relative to the main-text convention,
\begin{equation}
 \frac{\dd r_*}{\dd\rho}=\frac{\rho^2}{\rho^4-r_h^4},
 \qquad
 r_*-r_*^{(S)}=-\frac{\rho^3}{3r_h^4}+\cdots .
 \label{eq:appB-ads-tortoise}
\end{equation}
Hence the singular scaling again produces $q^2\omega^{-2/3}$ and $\omega^{-4/3}$.  Here the global Stokes network connects the horizon sector to the AdS-boundary sector through a clockwise half-turn of the local Bessel variable.  The infalling solution therefore has leading connection ratio
\begin{equation}
 S_0=-2\ii .
 \label{eq:appB-ads-S0}
\end{equation}
The finite-$q$ and temporal corrections carry precisely the two local grades above; the temporal Green integral requires the same WKB-overlap subtraction described in Appendix~\ref{app:framework}.

The crucial difference from asymptotically flat Schwarzschild is the outer boundary condition.  AdS normalizability fixes a definite ratio of the two WKB waves,
\begin{equation}
 B(\omega)=\ii e^{\ii\pi\nu}e^{-2\ii\omega x_0},
 \qquad
 x_0=\frac\beta4(1+\ii),\qquad \nu=\Delta_\Phi-2,
 \label{eq:appB-ads-boundaryratio}
\end{equation}
and the QNM condition is the open matching equation
\begin{equation}
 B(\omega_n)=S_{\rm QNM}(\omega_n).
 \label{eq:appB-ads-openmatch}
\end{equation}
At leading order this gives
\begin{equation}
 \omega_n^{(0)}=\frac{2\pi}{\beta}(1-\ii)\cN,
 \qquad
 \cN=n+\frac{\Delta_\Phi-3}{2}+\frac{\ii\log2}{2\pi},
 \label{eq:appB-ads-lattice}
\end{equation}
and root perturbation produces the $\cN^{-2/3}$ and $\cN^{-4/3}$ terms in Eq.~\eqref{eq:adsspectrum}.  The pure-AdS boundary expansion starts parametrically later and therefore cannot generate either leading fractional power independently.

\subsection{Gibbons--Maeda family}
\label{app:global-gm}

For rational $A=P/Q$, define $m=P+Q$, $d=P-Q$ and $H=1-b$, and uniformize the radial surface by
\begin{equation}
 u^m=1-\frac br,
 \qquad
 \frac{\dd r_*}{\dd u}
 =\frac{b^2m\,u^{2P-1}}{(u^m-H)(1-u^m)^2}.
 \label{eq:appB-gm-uniformizer}
\end{equation}
The physical singularity, event horizon and infinity are $u=0$, $u=c=H^{1/m}$ and $u=1$.  The horizon residue gives $\beta_H=4\pi c^d$, while the difference between the logarithmic period at infinity and that of the neighboring horizon sheet is
\begin{equation}
 \beta_I=2\pi\left(1-b\frac{d}{m}-c^d\right).
 \label{eq:appB-gm-betaI}
\end{equation}
The curvature singularity itself carries no logarithmic tortoise residue.

\begin{figure}[t]
 \centering
 \includegraphics[width=0.72\linewidth]{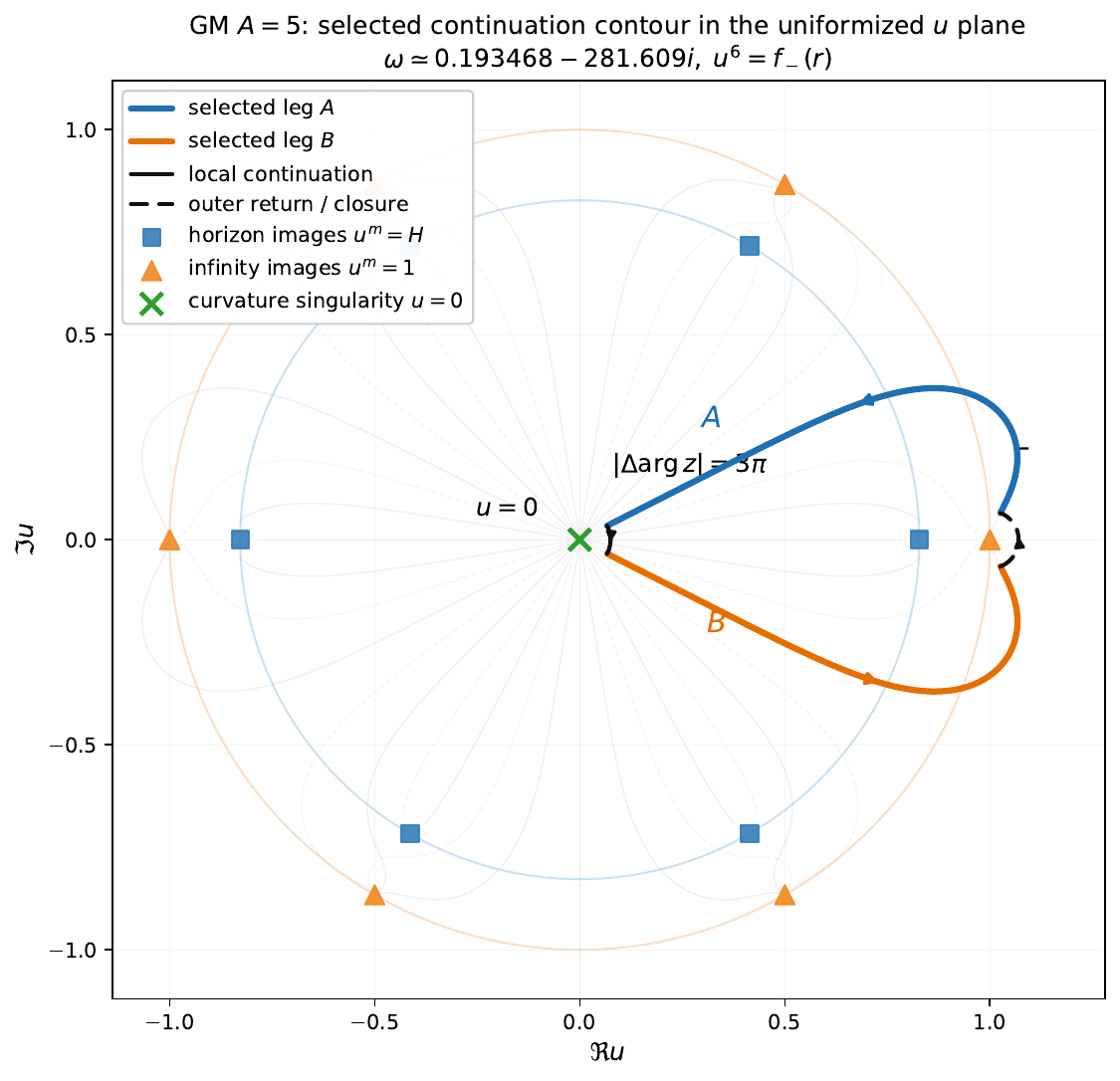}
 \caption{Representative complex-WKB continuation contour for the Gibbons--Maeda geometry at $A=5$ in the uniformized $u$ plane.  The blue and orange curves are the two physical-infinity-connected oscillatory legs, the short black segment denotes the local Bessel continuation through the singular region, and the dashed curve closes the contour through the physical-infinity neighborhood.  The figure labels the magnitude $|\Delta\arg z|=3\pi$; the retarded contour used in the analytic calculation is oriented clockwise, so that $\Delta\arg z=-3\pi$.  Squares and triangles denote the images of the horizon and infinity, respectively, and the green cross is the curvature singularity.}
 \label{fig:appB-gm-a5-stokes}
\end{figure}

Figure~\ref{fig:appB-gm-a5-stokes} shows a representative $A=5$ contour in the uniformized plane.  The two highlighted infinity-connected oscillatory legs enter the singular region on opposite sides of the physical continuation.  Their local separation has $|\Delta\arg z|=3\pi$; with the retarded orientation used throughout the paper the inner continuation is clockwise, $\Delta\arg z=-3\pi$.  Closing the two legs through the physical-infinity neighborhood produces a contour homologous to one positive winding around the physical event horizon and no winding around the remaining horizon or infinity images.

\begin{figure}[t]
 \centering
 \includegraphics[width=1\linewidth]{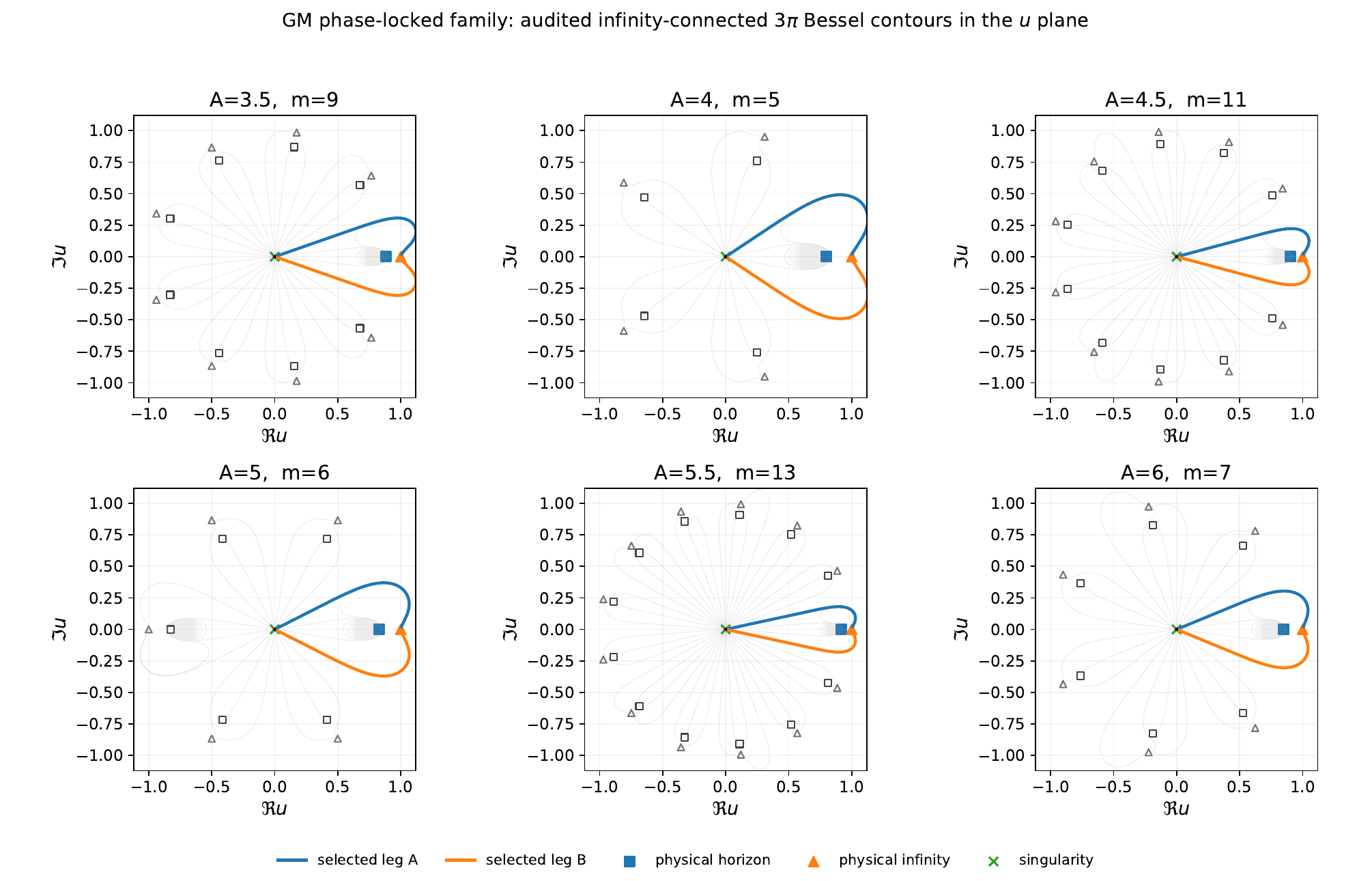}
 \caption{Family-wide Stokes audit for the six Gibbons--Maeda production geometries.  Faint curves show the full local oscillatory network launched from the curvature singularity, while the blue and orange curves are the two branches selected by their connection to the physical infinity.  The blue square, orange triangle and green cross denote the physical event horizon, physical infinity and curvature singularity.  Across all six values of $A$, the selected pair realizes the same $3\pi$ Bessel continuation and the same physical-horizon winding class.}
 \label{fig:appB-gm-six-stokes}
\end{figure}

The same topology is not special to the representative point.  Figure~\ref{fig:appB-gm-six-stokes} displays the corresponding infinity-connected pair for all six production geometries, $A=3.5,4,4.5,5,5.5,6$.  In every case the selected pair has $|\Delta\arg z|=3\pi$ and the associated closed contour winds the physical event horizon once while avoiding the other finite images.  This family-wide audit fixes the global contour used in the GM monodromy condition rather than assuming it from the $A=5$ example alone.

With $x=r-b$ and the singular-oriented coordinate $s$,
\begin{equation}
 s=\kappa_s x^{2A/(A+1)}[1+\Order(x)],
 \qquad
 V_0=-\frac{1}{4s^2}+\lambda_1s^{-2+\dO}+\cdots,
 \label{eq:appB-gm-local}
\end{equation}
while
\begin{equation}
 V_\Lambda=\Lambda\lambda_\Omega s^{-2+\dO}+\cdots,
 \qquad
 V_\mu=\mu^2\lambda_t s^{-2+\dt}+\cdots,
 \qquad
 \dO=\frac{A+1}{2A},\quad \dt=\frac{3A+1}{2A}.
 \label{eq:appB-gm-responses}
\end{equation}
A local insertion $\lambda_\rho s^{-2+\rho}$ therefore gives the leading response coefficient
\begin{equation}
 \mathcal C_\rho
 =-\frac23\lambda_\rho I(\rho)\sin(\pi\rho)
 \kappa_H^{1-\rho}e^{\ii\pi\rho/2}.
 \label{eq:appB-gm-Crho}
\end{equation}

Two further global facts enter the production fits.  First, the candidate massless $N^{-1}$ layer vanishes exactly.  In the uniformized plane the first diagonal WKB one-form is globally exact, $V_0\dd s=\dd F(u)$, and hence
\begin{equation}
 \Delta\Theta_\Gamma^{(1)}=-\frac{1}{2\omega}\oint_\Gamma V_0\,\dd s=0,
 \qquad B_{N^{-1}}=0.
 \label{eq:appB-gm-Bzero}
\end{equation}
Second, the analytically continued equation contains a distinct remote Stokes sector at $u=\infty$,
\begin{equation}
 R_{\rm rem}(\omega)
 =-4\cos\!\left[\frac{\pi(A+1)}{2(A+3)}\right]
 \exp\!\left\{\ii\beta_I\omega
 \left[\cot\!\left(\frac{\pi(A-1)}{A+1}\right)+\ii\right]\right\}.
 \label{eq:appB-gm-remote}
\end{equation}
This contribution is exponential in $\omega$ rather than part of the algebraic Puiseux series, and it has no explicit leading $\Lambda$ dependence.  It therefore produces the finite-overtone oscillatory nuisance in the raw massless spectrum but drops out of the numerator of the angular response.  In particular, changing $\beta_I/\beta_H$ changes the oscillation period without changing the local exponent $\dO$.

\section{Numerical solvers and robustness}
\label{app:numerics}

All quoted exponents are extracted from the full radial ODE, independently of the local connection calculation.  The three solvers differ, but in every case parameter responses are obtained by differentiating the spectral condition rather than subtracting nearby roots.

\subsection{Schwarzschild--AdS$_5$: Horowitz--Hubeny expansion}
\label{app:num-ads}

For $m^2=0$, $r_h=1$, ingoing Eddington--Finkelstein time and $z=1/r$ give
\begin{equation}
 z^2(1-z^4)\phi''+(-3z-z^5+2\ii\omega z^2)\phi'
 +(-3\ii\omega z-q^2z^2)\phi=0.
 \label{eq:appC-HH-ode}
\end{equation}
Expanding $\phi=\sum_{n\ge0}a_n(1-z)^n$ produces a six-term Horowitz--Hubeny recurrence.  At truncation $M$, let $F_M(\omega,Q)=\sum_{n=0}^M a_n$ with $Q=q^2$; QNMs satisfy $F_M=0$ and the clean momentum response is
\begin{equation}
 \left.\frac{\partial\omega_n}{\partial q^2}\right|_{0}
 =-\left.\frac{\partial_QF_M}{\partial_\omega F_M}\right|_{(\omega_n,0)}.
 \label{eq:appC-HH-response}
\end{equation}
The scaled complex response converges to the analytic $N^{-2/3}$ coefficient, while the massless frequency remainder converges independently to the $N^{-4/3}$ temporal coefficient; the blind effective exponents already approach $2/3$ and $4/3$ over $n=5$--$40$.

\subsection{5D Schwarzschild: Leaver--Nollert method}
\label{app:num-sch5}

After factoring the exact horizon-infalling and infinity-outgoing behaviors, the radial series obeys a four-term recurrence
\begin{equation}
 \alpha_na_{n+1}+\beta_na_n+\gamma_na_{n-1}+\delta_na_{n-2}=0,
 \label{eq:appC-LN-rec}
\end{equation}
which is reduced by Gaussian elimination to a three-term minimal-solution continued fraction.  The Nollert tail is expanded as
\begin{equation}
 \frac{a_{K+1}}{a_K}
 =1+c_1K^{-1/2}+c_2K^{-1}+\cdots,
 \qquad c_1^2=-2\ii k,\quad \Re c_1<0,
 \label{eq:appC-Nollert-tail}
\end{equation}
with the production runs retaining terms through $c_6$ and using $K=22000$.  If $F(\omega,L^2,\mu^2)=0$ is the continued-fraction equation,
\begin{equation}
 \left.
 \frac{\partial\omega_n}{\partial (L^2)}
 \right|_0
 =
 -\frac{F_{L^2}}{F_\omega},
 \qquad
 \left.
 \frac{\partial\omega_n}{\partial\mu^2}
 \right|_0
 =
 -\frac{F_{\mu^2}}{F_\omega}.
 \label{eq:appC-LN-responses}
\end{equation}
The $n=300$--$500$ sequences converge simultaneously in exponent and complex coefficient to the analytic $N^{-2/3}$ and $N^{-4/3}$ predictions.  The high-order Nollert tail is essential only for stabilizing the mass response at the highest overtones.

\subsection{Gibbons--Maeda: finite-band Evans formulation}
\label{app:num-gm}

For rational $A=P/Q$, uniformization and endpoint factoring convert the radial equation into a finite-band recurrence.  If the forward recurrence is written $a_{n+1}=-S_n/c_0(n)$, parameter sensitivities are propagated together with the coefficients,
\begin{equation}
 \partial_q a_{n+1}
 =-\frac{\partial_q S_n}{c_0(n)}
 +\frac{S_n\,\partial_q c_0(n)}{c_0(n)^2}.
 \label{eq:appC-GM-sensitivity}
\end{equation}
With the Birkhoff minimal tail $R_N$, an Evans function is
\begin{equation}
 E(\omega,q)=\frac{a_{N+1}}{a_N}-R_N,
 \qquad
 \left.\frac{\partial\omega_n}{\partial q}\right|_0=-\frac{E_q}{E_\omega},
 \label{eq:appC-GM-Evans}
\end{equation}
where $q=\Lambda$ gives the angular response and $q=\mu^2$ the temporal response.

The production raw-spectrum fit uses every consecutive overtone in the common window $n=80,\ldots,120$ for all six values $A=3.5,4,4.5,5,5.5,6$.  It fixes the analytically vanishing $N^{-1}$ term to zero, retains the local Puiseux hierarchy, and includes Eq.~\eqref{eq:appB-gm-remote} explicitly.  The angular response is fitted as
\begin{equation}
 S_n^{(\Omega)}
 =\mathcal C_\Omega(p)\mathcal N_n^{-p}
 +B_\Omega\mathcal N_n^{-2p},
 \qquad
 \mathcal N_n=n+\frac12+\frac{\ii\log3}{2\pi},
 \label{eq:appC-GM-angularfit}
\end{equation}
with $p$ and the subleading complex amplitude free; $\mathcal C_\Omega(p)$ is the independently derived coefficient in Eq.~\eqref{eq:appB-gm-Crho}.  The mass response uses the analogous leading coefficient and the measured angular grade for its subleading hierarchy.  No exact Kasner exponent or Kasner relation is imposed in either fit.

Solver and fit systematics are checked separately.  The production roots use a working-precision ladder and an $(N,K)$ recurrence-cutoff ladder; high-cutoff anchors reach $(N,K)=(1300,11)$, while independent long-window spectra at $A=3.5,4,5$ extend to $n=160$.  Increasing working precision at fixed cutoff changes the response far below the finite-cutoff drift, and the latter is included in the assigned numerical uncertainty.  These checks underlie the error bars shown in Fig.~\ref{fig:gmresults-a} and keep solver error distinct from finite-window fitting uncertainty.

The Fourier convention throughout is $e^{-\ii\omega t}$ with $\Im\omega<0$.  For the five-dimensional Schwarzschild and Gibbons--Maeda closed contours the retarded local rotation is $-3\pi$; for the AdS benchmark the boundary-connected sector lies on the lower lip of the negative local Bessel axis.

\bibliographystyle{JHEP}
\bibliography{references_jhep}

\end{document}